\documentclass[letterpaper]{jpconf}
\usepackage{graphicx}
\usepackage[left]{lineno}
\input{babarsym.tex}

\begin{document}
% \begingroup
% \linenumbers
\title{Measurement of the angle $\alpha$ at {\babar}}

\author{Alejandro P\'erez}

\address{Laboratoire de l'Acc\'el\'erateur Lin\'eaire, IN2P3/CNRS et Universit\'e Paris XI \\
         B\^atiment 200, 91898 Orsay cedex, France  \\
         On behalf of the {\babar} collaboration}

\begin{abstract}
We present recent measurements of the CKM angle $\alpha$ using data collected by the {\babar} 
detector at the PEP-II asymmetric-energy $e^+e^-$ collider at the SLAC National Acce- 
lerator Laboratory, operating at the $\Upsilon(4S)$ resonance. We present constraints on $\alpha$ 
from $B \to \pi\pi$, $B \to \rho\rho$ and $B \to \rho\pi$ decays.
% ~\\
% ~\\
% {\it Keywords:} time-dependent-asymmetry; CKM angle; longitudinal polarization.
\end{abstract}

\vspace{-1.2cm}
\section{Introduction}

The measurements of the angles $\alpha$, $\beta$ and $\gamma$ of the Unitarity Triangle (UT) 
at the B-factories are providing precision tests of the Standard Model (SM) description of $CP$ 
violation. This description is provided by the Cabibbo-Kobayashi-Maskawa (CKM) 
quark-mixing~\cite{CKMmatrix,Wolf}. We summarize the experimental constraints on the $\alpha$ 
UT angle obtained from $B$-meson  decays to $\pi\pi$, $\rho\rho$ and $\rho\pi$ with the {\babar} 
experiment at the SLAC National Accelerator Laboratory. The {\babar} detector and the PEP-II 
accelerator are described elsewhere~\cite{BaBar}.

\section{Analysis Method}

\subsection{General formula}

The decay of a neutral $B$-meson into a pair of $\pi$ or $\rho$ mesons, $B \to hh$ ($h = \pi,\rho$), occurs via 
two topologies: a tree-level process and a one-loop penguin diagram. The $CP$ parameter $\lambda_{hh}$, 
defined by $\lambda_{hh} = \frac{p}{q}\frac{\overline{A}}{A}$, where $q$ and $p$ are the complex coefficient 
that link the mass and the flavor eigenstates in the $B$ system, and $A$ ($\overline{A}$) is the $B^0$ 
($\overline{B}^0$) decay amplitude, can be expressed in terms of $\alpha$ as

\begin{equation}
\label{lambda_1}
\lambda_{hh} = e^{2i\alpha}
               \frac{1-(|V^*_{td}V_{tb}|/|V^*_{ud}V_{ub}|)P/Te^{-i\alpha}}
                    {1-(|V^*_{td}V_{tb}|/|V^*_{ud}V_{ub}|)P/Te^{i\alpha}}~,
\end{equation}
where $T$ and $P$ are complex amplitudes dominated by tree and penguin topologies, respectively.

The quantity experimentally measured is the time-dependent decay rate

\begin{equation}
\label{TDdecayAmp_1}
f_{Q_{\rm tag}} = \frac{e^{-|\Delta t|/\tau}}{4\tau}
                  \left[
                  1 - 
                  Q_{\rm tag}C_{hh}\cos(\Delta m_d\Delta t) + 
                  Q_{\rm tag}S_{hh}\sin(\Delta m_d\Delta t)
                  \right]~,
\end{equation}
where $\tau$ is the neutral $B$ lifetime and $\Delta m_d$ is the $B^0\overline{B}^0$ oscillation frequency. 
$\Delta t$ is the proper time difference between decays of the $B$ to $hh$ ($B_{\rm rec}$), and 
the second $B$ in the event, denoted by $B_{\rm tag}$. The $Q_{\rm tag}$ parameter is related to the 
flavor of the $B_{tag}$: $Q_{\rm tag} = +1 (-1)$ if the $B_{\rm tag}$ is a $B^0$ ($\overline{B}^0$). The 
$CP$-violating asymmetries $C_{hh}$ and $S_{hh}$ are related to the $\lambda_{hh}$ parameter by

\begin{equation}
\label{SandC}
S_{hh} = 2{\mathcal Im}(\lambda_{hh})/(1 + |\lambda_{hh}|^2)~, 
~~~~~~~~
C_{hh} = (1 - |\lambda_{hh}|^2)/(1 + |\lambda_{hh}|^2)~.
\end{equation}
$S_{hh}$ reflects the $CP$-violation induced by the interference between the mixing and decay processes; 
$C_{hh}$ is the direct $CP$-violating asymmetry which comes from the interference between different decay 
topologies. In the absence of penguin contributions ($P = 0$), $C_{hh}$ vanishes and $S_{hh}$ is simply 
related to the CKM angle $\alpha$ by $S_{hh} = \sin(2\alpha)$.

In the more general case of the $B^0(\overline{B}^0) \to \rho^{\pm}\pi^{\mp}$ decays, the time-dependent decay rate is 
given by
\begin{equation}
\label{TDdecayAmp_2}
f^{\rho^{\pm}\pi^{\mp}}_{Q_{\rm tag}} = (1 \pm {\mathcal A}_{\rho\pi})\frac{e^{-|\Delta t|/\tau}}{4\tau}
                  \left[1 -  Q_{\rm tag}(C_{\rho\pi} \pm \Delta C_{\rho\pi})\cos(\Delta m_d\Delta t)
                     +  Q_{\rm tag}(S_{\rho\pi} \pm \Delta S_{\rho\pi})\sin(\Delta m_d\Delta t)\right]~,
\end{equation}
where, the $\pm$ sign depends on whether the $\rho$ meson is emitted by the $W$ boson or comes from the spectator 
quark. ${\mathcal A}_{\rho\pi}$ is the direct $CP$ violation parameter measuring the asymmetry between the 
$\rho^+\pi^-$ and $\rho^-\pi^+$ final states, while $\Delta S_{\rho\pi}$ and $\Delta C_{\rho\pi}$, which arise from the fact 
that two production modes of the $\rho$ are possible, are dilution terms and have no $CP$ content.

\subsection{The isospin analysis}

Using the strong isospin symmetry, the angle $\alpha$ can be extracted up to discrete ambiguities from 
the $CP$-violating asymmetries defined above~\cite{LipkinNirQuinSnyder}. The decay amplitudes of the 
isospin-related final states obey the pentagonal relations

\begin{equation}
\label{isospin_1}
\sqrt{2}(A^{+0}_{\rho\pi} + A^{0+}_{\rho\pi})             = 2A^{00}_{\rho\pi} + 
                                                             A^{+-}_{\rho\pi} + 
                                                             A^{-+}_{\rho\pi}~,
~~~~~~~
\sqrt{2}(\overline{A}^{+0}_{\rho\pi} + \overline{A}^{0+}_{\rho\pi}) = 2\overline{A}^{00}_{\rho\pi} + 
                                                             \overline{A}^{+-}_{\rho\pi} + 
                                                             \overline{A}^{-+}_{\rho\pi}~;
\end{equation}
where $A^{ij}_{\rho\pi} = A(B^0~{\rm or}~B^+ \to \rho^i\pi^j)$ and 
$\overline{A}^{ij}_{\rho\pi} = A(\overline{B}^0~{\rm or}~B^- \to \rho^i\pi^j)$, $i,j = +,-,0$. With the use 
of these relations, 12 unknowns (6 complex amplitudes with one unphysical phase, and the CKM angle $\alpha$) are 
to be determined while 13 observables are available: $S_{\rho\pi}$, $C_{\rho\pi}$, $\Delta S_{\rho\pi}$, 
$\Delta C_{\rho\pi}$, ${\mathcal A_{\rho\pi}}$; four average branching fractions ${\mathcal B}(B \to \rho\pi)$; 
two time-dependent $CP$-violating asymmetries in the $B^0 \to \rho^0\pi^0$ decay ($S^{00}_{\rho\pi}$, 
$C^{00}_{\rho\pi}$) and two direct $CP$ asymmetries in $B^+ \to \rho^+\pi^0$ and $B^+ \to \rho^0\pi^+$ decays.

In the case of $B \to hh$ ($h=\pi,\rho$), Eq.~\ref{isospin_1} simplify to triangular relations
\begin{equation}
\label{isospin_2}
\sqrt{2}A^{+0}_{hh} = A^{+-}_{hh}        + A^{00}_{hh}~,
~~~~~~~
\sqrt{2}\overline{A}^{+0}_{hh} = 2\overline{A}^{+-}_{hh} + \overline{A}^{00}_{hh}~.
\end{equation}
The information counting leads then to 6 unknowns and 7 observables: 3 branching fractions 
${\mathcal B}(B\to hh)$; $C_{hh}$, $S_{hh}$, $C^{00}_{hh}$, $S^{00}_{hh}$. In the $\pi\pi$ system 
$S^{00}_{\pi\pi}$ is impossible to measure (as the $\pi^0$ is reconstructed from two-photons decays, 
there is no way to measure the decay vertex), then one is left with 6 observables: $\alpha$ can 
be extracted with an 8-fold ambiguity within $[0,\pi]$~\cite{GronauLondon}.

\section{Experimental Results}

\subsection{$B \to \pi\pi$ and $B \to \rho\rho$}

The various branching fractions and $CP$-asymmetries measured in $B \to \pi\pi$ and $B \to \rho\rho$ 
decays are summarized in Table~\ref{tab:pipi_rhorho}. In the case of charged decays the charge asymmetry 
is defined as ${\mathcal A}_{CP}(B \to hh) = -C_{hh}$. The measurements are sufficiently well established 
to perform an isospin analysis.

\begin{table}[hbt!]
\begin{center}
\begin{TableSize}
\begin{tabular}{cccc}
\hline
{\bf Mode}         & ${\mathbf {\mathcal B}(10^{-6})}$ & ${\mathbf C}$                    & ${\mathbf S}$ \\
\hline
$\pi^+\pi^-$       & $5.5 \pm 0.4 \pm 0.3$~\cite{BF_BTopippim} 
                   & $-0.68 \pm 0.10 \pm 0.03$~\cite{LattestBTopipi} 
                   & $-0.25 \pm 0.08 \pm 0.02$~\cite{LattestBTopipi} \\
$\pi^0\pi^0$       & $1.83 \pm 0.21 \pm 0.13$~\cite{LattestBTopipi} 
                   & $-0.43 \pm 0.26 \pm 0.05$~\cite{LattestBTopipi} 
                   & -- \\
\hline
$\rho^+\rho^-$     & $25.5 \pm 2.1^{+3.6}_{-3.9}$~\cite{BTorhoprhom} 
                   & $0.01 \pm 0.15 \pm 0.06$~\cite{BTorhoprhom} 
                   & $-0.17 \pm 0.2^{+0.05}_{-0.06}$~\cite{BTorhoprhom} \\
$\rho^0\rho^0$     & $0.92 \pm 0.32 \pm 0.14$~\cite{BTorho0rho0} 
                   & $0.2 \pm 0.8 \pm 0.3$~\cite{BTorho0rho0} 
                   & $0.3 \pm 0.7 \pm 0.2$~\cite{BTorho0rho0} \\
\hline
{\bf Mode}         & ${\mathbf {\mathcal B}(10^{-6})}$ & ${\mathbf {\mathcal A_{CP}}}$ &  \\
\hline
$\pi^{\pm}\pi^0$   & $5.02 \pm 0.46 \pm 0.29$~\cite{BF_BTopippi0} 
                   & $0.03 \pm 0.08 \pm 0.01$~\cite{BF_BTopippi0} 
                   & \\
\hline
$\rho^{\pm}\rho^0$ & $23.7 \pm 1.4 \pm 1.4$~\cite{BTorhoprho0} 
                   & $-0.054 \pm 0.055 \pm 0.010$~\cite{BTorhoprho0} 
                   & \\
\hline
\end{tabular}
\end{TableSize}
\end{center}
\caption{\em Summary of {\babar} measurements of $B \to \pi\pi$ and $B \to \rho\rho$ decays. 
The measurements for the $\rho\rho$ system corresponds to the longitudinal component of the decay 
rate. The errors quoted are statistical and systematic, respectively.
\label{tab:pipi_rhorho}}
\end{table}

The present measurement for the $\pi^+\pi^-$ mode excludes the absence of $CP$ violation 
$(C_{\pi\pi},S_{\pi\pi}) = (0,0)$ at a C.L. of $6.7\sigma$. The relatively high branching 
fraction of the $\pi^0\pi^0$ mode tends to separate the 8-fold ambiguities in the 
$\alpha$ extraction, which only allows a weak constraint on $\alpha$ to be set.
With the current experimental measurements two of the eight ambiguities are nearly merged. 
The range $[23^o,67^o]$ in $\alpha$ is excluded at the $90\%$ C.L.~\cite{LattestBTopipi}. 
The solution is in agreement with the global CKM fit~\cite{CKMfitter,UTfit} which gives the 
range $[71^o,109^o]$ at $68\%$ C.L.

The analysis of $B \to \rho\rho$ is potentially complicated due to the possible presence 
of three helicity states for the decay. The helicity zero state, which corresponds to longitudinal 
polarization of the decay, is $CP$-even but the helicity $\pm 1$ states are not $CP$ eigenstates. 
Fortunately this complication is avoided by the experimental finding that the dominant polarization 
is longitudinal, $f_L(\rho^+\rho^-) = 0.992 \pm 0.024^{+0.026}_{-0.013}$~\cite{BTorhoprhom}, 
$f_L(\rho^0\rho^0) = 0.75^{+0.11}_{-0.14} \pm 0.05$~\cite{BTorho0rho0} and 
$f_L(\rho^+\rho^0) = 0.950 \pm 0.015 \pm 0.006$~\cite{BTorhoprho0} ($f_L \equiv \Gamma_L/\Gamma$, 
where $\Gamma$ is the total decay rate and $\Gamma_L$ is the rate of the longitudinally-polarized mode). 
The $B^0 \to \rho^0\rho^0$ branching fraction is small compared with that of the $B^+ \to \rho^+\rho^0$ 
mode, which indicates that the penguin to three ratio ($P/T$, cf. Eq.~\ref{lambda_1}) is small compared with 
that of the $B \to \pi\pi$ system~\cite{LipkinNirQuinSnyder}. This has the effect of merging the 
different ambiguities in the extraction of $\alpha$. The latest $B^0 \to \rho^0\rho^0$ {\babar} 
results present the first measurement of the time-dependent $CP$ asymmetries $C^{00}_L$ and $S^{00}_L$. The inclusion of 
these measurements has the effect of raising the 8-fold degeneracy on $\alpha$: the data only favors 
two solutions out of eight~\cite{BTorho0rho0,BTorhoprho0}. These two effects allow to set a strong 
constraint on $\alpha$, where only two solutions are seen, corresponding to 
$\alpha = (92.4^{+6.0}_{-6.5})^o$ at $68\%$ C.L.~\cite{BTorhoprho0} for the one in agreement with 
the global CKM fit~\cite{CKMfitter,UTfit}.

\subsection{$B \to \rho\pi$}

The $B \to \rho\pi$ measurement reported here is a time-dependent amplitude analysis of $B^0 \to (\rho\pi)^0$. 
The interferences between the intersecting $\rho$ resonance bands are modeled over the whole Dalitz Plot 
using the isobar model~\cite{IsobarModel}. This allows determination of the strong phase differences from the 
interference pattern, which permits direct extraction of the angle $\alpha$ with reduced ambiguities. 
The Dalitz amplitudes and time-dependence are contained in the 26 coefficients of the bilinear form-factor 
terms occurring in the time-dependent decay rate, which are determined from a likelihood fit. 
The values obtained for these coefficients are converted back into the quasi-two-body $CP$ observables 
(c.f. Eq.~\ref{TDdecayAmp_2}), which are more intuitive in their interpretation. Table~\ref{tab:rhopi} 
reports the experimental findings on these observables~\cite{RhoPi}.

\begin{table}[hbt!]
\begin{center}
\begin{TableSize}
\begin{tabular}{lc|lc}
\hline
{\bf Observable}           & {\bf Value}                  & {\bf Observable}          & {\bf Value} \\
\hline
$C_{\rho\pi}$              & $ 0.15 \pm 0.09 \pm 0.05$    & $S_{\rho\pi}$             & $-0.03 \pm 0.11 \pm 0.04$ \\
$\Delta C_{\rho\pi}$       & $ 0.39 \pm 0.09 \pm 0.09$    & $\Delta S_{\rho\pi}$      & $-0.01 \pm 0.14 \pm 0.06$ \\
$C^{00}_{\rho\pi}$         & $-0.10 \pm 0.40 \pm 0.53$    & $S^{00}_{\rho\pi}$        & $ 0.04 \pm 0.44 \pm 0.18$ \\
$A_{\rho\pi}$              & $-0.14 \pm 0.05 \pm 0.02$    &                           &                           \\
\hline
\end{tabular}
\end{TableSize}
\end{center}
\caption{\em Summary of {\babar} measurements from the time-dependent amplitude analysis of 
$B^0 \to (\rho\pi)^0$ decays. The errors quoted are statistical and systematic, respectively.
\label{tab:rhopi}}
\end{table}

These measurements allow the determination of the limit $\alpha = (87^{+45}_{-13})^o$ at $68\%$ C.L., 
with almost no constraint at $95\%$ C.L. This result is particularly interesting as there is an unique 
solution in the $[0,180]^o$ range, which helps to break the ambiguities obtained from the $\pi\pi$ and 
$\rho\rho$ results. A hint of $CP$-violation is obtained at the level of $3\sigma$.

\section{Summary}

\begin{figure}[h!]
\begin{center}
\begin{minipage}{.49\linewidth}
\hspace{0.5cm}
\vspace{-0.8cm}
\includegraphics[width=6.0cm,keepaspectratio]{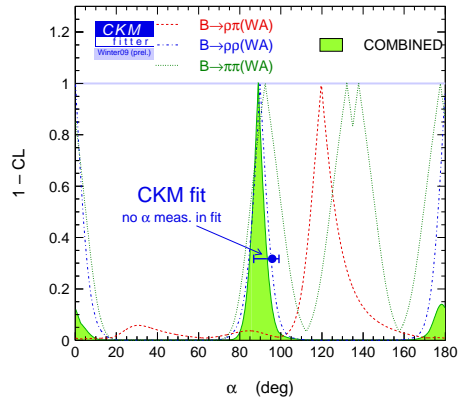}
\end{minipage}
\end{center}
\caption
{\label{fig:alpha_average_wa}
{\em Constraints on $\alpha$, provided by the CKMfitter group~\cite{CKMfitter}, expressed as one minus the 
confidence level as a function of angle. The constraints are constructed averaging the {\babar} and 
Belle measurements for the $\pi\pi$ (dotted green curve), $\rho\rho$ (dash-dotted blue curve) and $\rho\pi$ 
(dashed red curve) systems. The solid filled green curve represents the combined constraint using all the 
systems.
}}
\end{figure}

Several analyses have been conducted in {\babar} to extract the angle $\alpha$ of the UT. In the last few years 
the measurements of this angle have become increasingly precise. The measurements provided from the 
$B \to \pi\pi/\rho\rho/\rho\pi$ modes give complementary constraints on $\alpha$. For the $B \to \rho\rho$ system, 
the inclusion of the $S^{00}_{\rho\pi}$ observable allows to favor two of the 8-fold ambiguities on $\alpha$, and 
the relatively large ${\mathcal B}(B^+ \to \rho^+\pi^0)$, with respect to ${\mathcal B}(B^0 \to \rho^0\pi^0)$, 
causes the ambiguities to degenerate in two peaks, improving the precision of the constraint. The measurements 
from the $B^0 \to (\rho\pi)^0$ time-dependent amplitude analysis give a direct access to $\alpha$, disfavoring 
the ambiguities. The combined constraint averaging all the $\pi\pi$, $\rho\rho$ and $\rho\pi$ measurements from 
{\babar} and Belle gives $\alpha = (89.0^{+4.4}_{-4.2})^o$ at $68\%$ C.L. (see Fig.~\ref{fig:alpha_average_wa}), 
which is in good agreement with the global CKM fit~\cite{CKMfitter,UTfit}.

\vspace{-0.7cm}

\vspace{-0.8cm}

\section{Acknowledgements}

I would like to thank the organizers of the Lake Louise Winter Institute 2009 for an enjoyable and stimulating 
conference, and my {\babar} colleagues for their assistance and helpful discussions.

% \vspace{-0.3cm}

% \endgroup
\end{document}